\documentclass[twocolumn,aps,showpacs,prl,tightenlines,amsmath,amssymb,superscriptaddress]{revtex4}
\usepackage{graphicx}

\usepackage{dcolumn}

\usepackage{bm}

\usepackage{colordvi}

\begin{document}

\title{Effect of initial spin polarization on spin dephasing and electron $g$ factor in a high-mobility two-dimensional electron system}

\author{D.\ Stich}
\affiliation{Institut f\"ur Experimentelle und Angewandte Physik,
Universit\"at Regensburg, D-93040 Regensburg, Germany}
\author{J.\ Zhou}
\affiliation{Hefei National Laboratory for Physical Sciences at
Microscale and Department of Physics, University of Science and
Technology of China, Hefei, Anhui, 230026, China}
\author{T.\ Korn}
\affiliation{Institut f\"ur Experimentelle und Angewandte Physik, Universit\"at Regensburg,
D-93040 Regensburg, Germany}
\author{R.\ Schulz}
\affiliation{Institut f\"ur Experimentelle und Angewandte Physik,
Universit\"at Regensburg, D-93040 Regensburg, Germany}
\author{D.\ Schuh}
\affiliation{Institut f\"ur Experimentelle und Angewandte Physik,
Universit\"at Regensburg, D-93040 Regensburg, Germany}
\author{W.\ Wegscheider}
\affiliation{Institut f\"ur Experimentelle und Angewandte Physik,
Universit\"at Regensburg, D-93040 Regensburg, Germany}
\author{M.\ W.\ Wu}
\email{mwwu@ustc.edu.cn.}
\affiliation{Hefei National Laboratory for Physical Sciences at
Microscale and Department of Physics,
University of Science and Technology of China, Hefei,
Anhui, 230026, China}
\author{C.\ Sch\"uller}
\email{christian.schueller@physik.uni-regensburg.de.}
\affiliation{Institut f\"ur Experimentelle und Angewandte Physik,
Universit\"at Regensburg, D-93040 Regensburg, Germany}

\date{\today}

\begin{abstract}
We have investigated the spin dynamics of a high-mobility
two-dimensional electron system (2DES) in a
GaAs--Al$_{0.3}$Ga$_{0.7}$As single quantum well by time-resolved
Faraday rotation (TRFR) in dependence on the initial degree of
spin polarization, $P$, of the 2DES. From $P\sim 0$ to $P\sim
30$~\%, we observe an increase of the spin dephasing time,
$T_2^\ast$, by an order of magnitude, from about 20 ps to 200 ps,
in good agreement with theoretical predictions by Weng and Wu
[Phys. Rev. B {\bf 68}, 075312 (2003)]. Furthermore, by applying
an external magnetic field in the Voigt configuration, also the
electron $g$ factor is found to decrease for increasing $P$. Fully
microscopic calculations, by numerically solving the kinetic spin
Bloch equations considering the D'yakonov-Perel' and the
Bir-Aronov-Pikus mechanisms, reproduce the most salient features
of the experiments, {\em i.e}., a dramatic decrease of spin
dephasing and a moderate decrease of the electron $g$ factor with
increasing $P$. We show that both results are determined
dominantly by the Hartree-Fock contribution of the Coulomb
interaction.
\end{abstract}


\maketitle

The spin degrees of freedom in semiconductors have largely been
explored in recent years, motivated by potential applications in
spintronics or quantum computational devices
\cite{Awschalom1,Fabian}. Very naturally, the spin dephasing and
spin relaxation is of utmost interest in this field. For the
paradigm spintronics device, the so-called spin transistor, as
proposed by Datta and Das \cite{DattaDas}, injection and control of
spin-polarized electrons into a field-effect semiconductor
transistor structure is required. In a number of pioneering
experiments, Kikkawa and Awschalom {\em et al.}
\cite{Kikkawa1,Kikkawa2} have demonstrated that extraordinarily long
spin relaxation times can be achieved in GaAs bulk material
\cite{Kikkawa1} or in II-VI quantum wells \cite{Kikkawa2} by using
doping levels close to the metal-to-insulator transition. This
provides on the one hand an outstanding means for manipulation of
optically-excited spins. On the other hand, however, a high impurity
concentration seems to be contradictory for a transistor device,
where highly-mobile charge carriers are required for a minimum of
dissipation processes. Definitely, another pioneering step into this
direction was the experimental verification of a strongly reduced
spin dephasing in GaAs-based quantum well structures, which are
grown in the $[110]$ crystal direction by Ohno {\em et al.}
\cite{Ohno,Dohrmann}. Up to that point, the commonly used
theoretical considerations, employing the D'yakonov-Perel' (DP) spin
relaxation mechanism \cite{dp}, predicted the absence of spin
relaxation due to this mechanism for spins aligned parallel to this
particular crystal direction. However, it was pointed out by Wu and
Kuwata-Gonokami \cite{WuSSC} that in the presence of a magnetic
field in the Voigt configuration, there is still inhomogeneous
broadening which can result in a finite spin dephasing contribution
due to the DP mechanism. Moreover, in particular in high-mobility
two-dimensional electron systems (2DES), as favorable components of
field-effect transistor-like structures, electron-electron
interaction can play a major role. This is because, as first pointed
out by Wu and Ning \cite{Wu}, any scattering including the Coulomb
scattering can cause an irreversible spin dephasing in the presence
of inhomogeneous broadening. Recently, also for standard
$[001]$-oriented $n$-doped quantum wells, the importance of the
electron-electron scattering for spin relaxation and dephasing was
demonstrated by Glazov and Ivchenko \cite{Ivchenko} from
perturbation theory and Weng and Wu \cite{wu1} from a fully
microscopic many-body approach. In most of the previous theoretical
and experimental studies, the spin systems are near the equilibrium,
{\em i.e.}, the spin polarization is very small and there is
no/small external electric field parallel to the quantum wells.
Nevertheless, Weng and Wu investigated the spin kinetics far away
from the equilibrium \cite{wu1,wu2} by setting up and numerically
solving the kinetic spin Bloch equations \cite{wumetiu} with all the
scattering, especially the electron-electron Coulomb scattering
explicitly included. One of the predictions of the spin dephasing
far away from the equilibrium is that the spin dephasing is greatly
suppressed by increasing the initial spin polarization. This effect
originates from the Hartree-Fock (HF) contribution of the Coulomb
interaction which serves as an effective magnetic field along the
$z$-axis. This effective magnetic field is greatly increased with
the spin polarization and therefore blocks the spin precession due
to the lack of detuning \cite{wu1}.

In this Letter, we manage to realize the large spin polarization
experimentally and show that the proposed effect does exist.
Spin-polarized carriers are injected at the Fermi level of the 2DES
via optical pumping with circularly-polarized light. The ensemble
spin dephasing time $T_2^{\ast}$ is experimentally determined via
time-resolved Faraday rotation (TRFR). We find a strong increase of
$T_2^{\ast}$ with increasing initial spin polarization in good
agreement with the fully microscopic theory \cite{wu1}. Furthermore,
also the variation of the electron $g$ factor with degree of spin
polarization shows, both in experiment and theory, the same
tendency. This verifies and underlines the importance of
electron-electron interaction for electron spin dephasing in
high-mobility 2DES.

\begin{figure}[t]
\begin{center}\includegraphics[width=7.5cm]{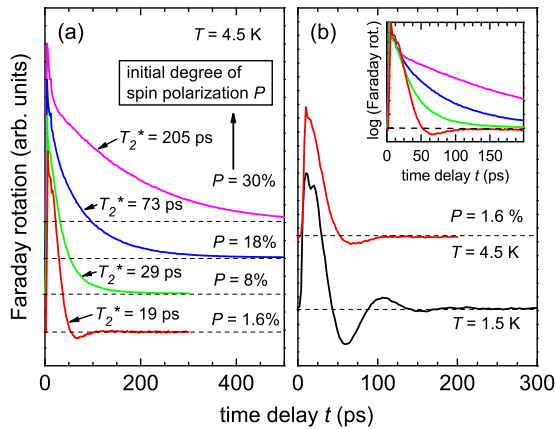}\end{center}
\caption{(color online) (a) Normalized TRFR traces for different
degrees of initial electron spin polarization, $P$. The total
densities of electrons, $n_{tot}=n_e+n_{ph}^{tot}$, are 2.19, 2.66,
3.83, 8.39 (in units of 10$^{11}$ cm$^{-2}$) for $P=1.6\% $, $8\%$,
$18\%$, $30\%$, respectively. (b) Comparison of two traces with
$P=1.6\%$ at two different temperatures. For the lowest temperature,
a coherent electron spin oscillation due to the Dresselhaus-Rashba
spin-orbit field is clearly seen. The inset shows a semilogarithmic
plot of the data displayed in (a). Note that at low $P$ the
zero-field oscillation is superimposed to the decay.}\label{Fig1}
\end{figure}

Our sample is a $[001]$-grown, 20 nm-wide, one-sided
modulation-doped GaAs-Al$_{0.3}$Ga$_{0.7}$As single quantum well.
The electron density and mobility at $T=4.2$ K are $n_e=2.1\times
10^{11}$ cm$^{-2}$ and $\mu_e=1.6\times 10^6$ cm$^2$/Vs,
respectively. For measurements in transmission geometry, the sample
was glued onto a glass substrate with an optical adhesive, and the
substrate and buffer layers were removed by selective etching. The
sample was mounted in the $^3$He insert of a superconducting
split-coil magnet cryostat. For the TRFR measurements, two laser
beams from a mode-locked Ti:Sapphire laser, which is operated at 80
MHz repetition rate, were used. The laser pulses had a temporal
length of about 600 fs each, resulting in a spectral width of about
3-4 meV, which allowed for a resonant excitation. The laser
wavelength was tuned to excite electrons from the valence band to
states slightly above the Fermi energy of the host electrons in the
conduction band. Both laser beams were focused to a spot of
approximately 30 $\mu$m diameter on the sample surface. The pump
pulses were circularly polarized in order to create spin-oriented
electrons in the conduction band, with spins aligned perpendicular
to the quantum well plane. Average pump powers between about 100
$\mu$W and 6 mW were used to create different densities, $n_{ph}$,
of photoexcited, spin-aligned electrons. We have estimated the total
densities, $n_{ph}^{tot}$, of electron-hole pairs, created by a pump
pulse to be between about $n_{ph}^{tot}=9\times 10^9$ cm$^{-2}$ for
the lowest, and $n_{ph}^{tot}=6\times 10^{11}$ cm$^{-2}$ for the
highest pump intensities\ \cite{densities}. Referring to $k\cdot p$
calculations of Pfalz {\em et al.}\ \cite{Pfalz}, we have determined
for our 20 nm-wide GaAs well the densities of spin aligned electrons
$n_{ph}$ by multiplying $n_{ph}^{tot}$ by a factor of 0.4 to account
for heavy-light hole mixing in the valence band. The resulting
degree of initial spin polarization of electrons was then calculated
via the relation $ P={n_{ph}}/{(n_e+n_{ph}^{tot})}$. The intensity
of the linearly polarized probe pulses was kept constant at an
average power of about 0.5 mW, and the rotation of the probe
polarization due to the Faraday effect was measured by an optical
bridge.

\begin{figure}[t]
\begin{center}\includegraphics[width=7.5cm]{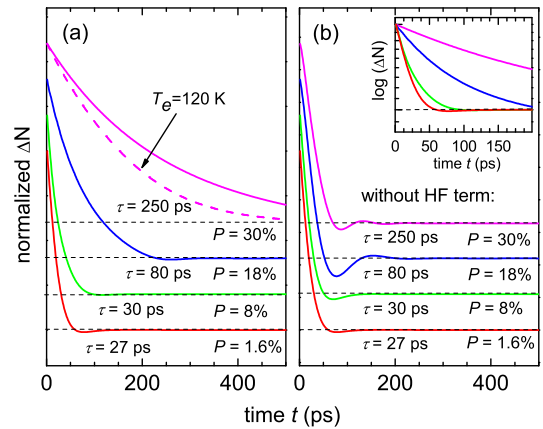}\end{center}
\caption{(color online) (a) Calculated spin decay curves for the
experimental parameters, like initial spin polarization, total
electron densities, electron mobility, and temperature $T=4.5$ K
(solid lines). In the calculation, a phenomenological decay time
 is incorporated as a single fitting
parameter $\tau$ (see text). The dashed curve
 for $P=30\%$ is calculated for a hot electron temperature
of $T_e=120$\ K. (b) Same as (a) but calculated without the HF term.
In particular for large $P$ the decay is much faster than in the
experiments (cf.~Fig.\ 1a). The inset shows the data, displayed in
(a), in a semilogarithmic plot. At low $P$ the zero-field
oscillations are superimposed. }\label{Fig2}
\end{figure}

Figure\ \ref{Fig1}a displays measured TRFR traces at a temperature
of $T=4.5$ K for different degrees of initial electron spin
polarization, $P$. As described above, $P$ was set by the average
pump powers applied in the pump beam. Thus, from bottom to top
spectra in Fig.\ \ref{Fig1}a, together with $P$, also the total
electron densities increase (numbers are given in the figure
caption) \cite{lifetime}. In particular at higher pump intensities,
a biexponential decay can be found. There is a very fast decay of
the TRFR signal within the first few picoseconds, followed by a
second, much longer decay. We attribute the fast initial decay to
the spin relaxation of the photoexcited holes, which lose their
initial spin orientation extremely fast. The second long decay is
due to electron spin relaxation. The extracted electron spin
dephasing times $T_2^{\ast}$, given in Fig.\ \ref{Fig1}a, increase
significantly with increasing $P$. For the lowest $P$ in Fig.\
\ref{Fig1}a, a very weak oscillation can be observed. As Fig.\
\ref{Fig1}b shows, this oscillation increases, when the temperature
is lowered from 4.5 K to 1.5\ K. A very similar zero-field spin
oscillation was previously reported by Brand {\em et al.}
\cite{Brand}. The oscillatory signal is due to a coherent
oscillation of the excited electron spins
 about an effective spin-orbit field  caused by $k$-linear terms in the
Rashba-Dresselhaus Hamiltonian. The inset of Fig.~\ref{Fig1}b shows
a semilogarithmic plot of the experimental curves, displayed in
Fig.~\ref{Fig1}a. Note that in particular at low $P$ the zero-field
oscillation is superimposed on the decay. At high $P$, the
exponential decay is clearly visible.

\begin{figure}[t]
\begin{center}\includegraphics[width=5.3cm]{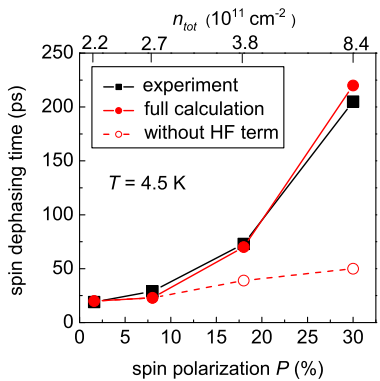}\end{center}
\caption{(color online) Comparison of spin dephasing times, as
extracted from the experimental curves in Fig.\ 1a (solid squares),
and from the calculated curves in Fig.\ 2a with (solid dots), and
Fig.\ 2b without (open circles) the HF term. In both calculations,
phenomenological decay times, $\tau$, as indicated in Fig.\ 2, were
included.} \label{Fig3}
\end{figure}

In order to understand the experimental results, we have performed a
fully microscopic calculation on the corresponding cases by setting
up and numerically solving the kinetic spin Bloch equations
\cite{wu1,wu2,wu3}. In the calculation we only consider the DP
mechanism by including the Dresselhaus term \cite{dress}. All the
spin-conserving scattering such as the electron-acoustic-phonon,
electron-nonmagnetic-impurity, electron-electron and
electron-heavy-hole Coulomb scatterings as well as spin-flip
electron-hole Coulomb scattering [the so called Bir-Aronov-Pikus
(BAP) mechanism \cite{bap}] are included. The impurity density is
determined from the Hall mobility measured in the experiment. The
details of the calculation can be found in Ref.\ \onlinecite{wu3},
except now the screening is expanded to include the contribution
from heavy holes. It is seen that the theoretical calculations,
displayed in Fig.\ \ref{Fig2}a, qualitatively reproduce the
experimental findings very well: The very weak zero-field spin
oscillation is observed for small $P$ and disappears for higher $P$.
Furthermore, the dramatic decrease of the spin dephasing with $P$,
first predicted by Weng and Wu in Ref.\ \onlinecite{wu1},
 is reproduced in the calculation. This quick decrease of the
spin dephasing with the
 spin polarization was suggested to be due to the HF contribution from the
Coulomb interaction \cite{wu1}. This can be seen clearly by
comparing the corresponding curves in Fig.\ \ref{Fig2}a with
Fig.\ \ref{Fig2}b where the HF term is removed.

It is seen that a fitting parameter $\tau$ is introduced in Fig.\ \ref{Fig2}
  so that we may obtain
the same $T_2^\ast$ as the experiment. This phenomenological
parameter is used to represent (i) additional sources of spin
dephasing, not included in the present model, such as the effect
from the asymmetric quantum well in the experiment (the Rashba
term), the multi-subband effect at large $P$, and, (ii) the fact
that with the finite laser pulse width of about 3-4\ meV in the
experiment, we cannot probe the complete energy range above the
Fermi energy, as we do in the calculations. It is seen that $\tau$
increases with increasing $P$: For high polarizations the
photoexcited hole density becomes much higher and consequently the
BAP mechanism becomes stronger. Therefore, the remaining unaccounted
mechanisms become less significant, and hence $\tau$ becomes larger.
It is noted that in Fig.\ \ref{Fig2} the {\em same} parameters
$\tau$ are used for {\em both} calculations with and without the HF
term. Moreover, as the actual temperature in the laser spot can be
higher than what we measure at the sample holder (4.5\ K), for the
highest pump power $P=30$~\%, we present another calculation by
taking hot electron temperature $T_e=120$\ K as dashed curve\
\cite{temperature}. It shows that, theoretically, the spin dephasing
time decreases with temperature, which is in opposite to the
temperature dependence at small spin polarization \cite{wu3}. This
indicates that the enhancement of the spin dephasing time with the
spin polarization is {\em solely} due to the HF term even if the
temperature is raised at high $P$. Finally, as a summary, in Fig.\
\ref{Fig3} the spin dephasing times, as extracted from the curves in
Figs.\ \ref{Fig1} and \ref{Fig2} are displayed versus $P$ and total
electron density $n_{tot}$. As mentioned above, a phenomenological
$\tau$ (see Fig.\ \ref{Fig2}) is included in the calculations to
derive the same dephasing times as in the experiment. We want to
emphasize that the important result here is the striking difference
of the calculations with (solid dots) and without the HF term (open
circles) for the {\em same} $\tau$ \cite{flat}.

\begin{figure}[tbh]
\begin{center}\includegraphics[width=7.6cm]{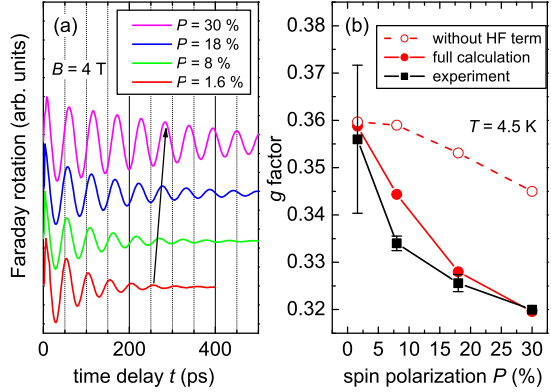}\end{center}
\caption{(color online) (a) TRFR measurements at $B=4$ T for
different $P$. An increase of the electron precession period with
increasing $P$ is clearly observed (arrow). (b) Comparison of
electron $g$ factors for different polarization degrees $P$, as
extracted from the experiments (solid squares), and the
calculations with (solid dots) and without (open dots) HF
term.}\label{Fig4}
\end{figure}

We now turn to the discussion of experiments in an external
magnetic field, applied in Voigt geometry, {\em i.e.}, parallel to
the quantum-well plane.  Figure\ \ref{Fig4}a shows
 TRFR traces
for fixed magnetic field of
$B=4$\ T but different spin polarizations $P$.
 One can clearly see that,
as $P$ is increased, the precession period increases (see arrow in
Fig.\ \ref{Fig4}a). We plot in Fig.\ \ref{Fig4}b the polarization
dependence of the electron $g$ factor which we have extracted from
the experimental curves. One can see that $g$ decreases with $P$ by
nearly $10$\ \% while $P$ is increasing from $1.6$~\% to $30$~\%.
The calculations with and without the HF term also indicate the
marked variation of $g$, with the variation for the former case
being larger than the latter one. This shows that in the weak
scattering limit, both the effective magnetic fields from the
Dresselhaus term and the HF term change the frequency of Larmor
precession.

In conclusion, we have investigated the spin dephasing and the
electron $g$ factor of a high-mobility 2DES in dependence of an
initial electron spin polarization (especially far away from the
equilibrium).  Experimentally, the initial spin polarization was
realized by optically pumping spin-aligned electrons into states
above the Fermi energy of the 2DES. We have found a marked increase
of the spin dephasing time with spin polarization, and a slight
decrease of the electron $g$ factor. The experimental results could
be confirmed by fully microscopic calculations, employing the
kinetic spin Bloch equations. Importantly, it turned out that in the
calculations the HF contribution from the Coulomb interaction plays
a dominant role and is responsible for the dramatic increase of the
spin dephasing time at large spin polarization. It can by no means
be neglected in order to correctly account for the experimental
observations.

We gratefully thank Jaroslav Fabian and R. T. Harley for valuable
discussions. This work was supported by the Deutsche
Forschungsgemeinschaft via GrK 638, grant No. Schu1171/1-3, SFB 689,
and SPP1285. M.W.W. was supported by the Natural Science Foundation
of China under Grant No.\ 10574120, the National Basic Research
Program of China under Grant No.\  2006CB922005, the Knowledge
Innovation Project of Chinese Academy of Sciences and SRFDP.


\begin{thebibliography}{0}
\bibitem{densities} For estimation of the densities of photoexcited
electron-hole pairs, $n_{tot}^{ph}$, the energy-dependent
absorption coefficient of the sample, and the laser spot size were
measured. From this, and the laser beam intensity, $n_{tot}^{ph}$
was calculated.
\bibitem{lifetime}The measured photocarrier lifetime (not shown
here) was on the order of 1 ns so that we assume that lifetime
effects on the spin relaxation are of minor importance.
\bibitem{temperature} We have estimated the increase of electron
temperature with laser pump power by analyzing the
photoluminescence lineshape at similar excitation conditions. From
this we get a moderate increase from about $T_e=7 $ K to $T_e=18$
K for $P=1.6 $\ \% to $P=30$\ \%.
\bibitem{flat}Moreover, without
$\tau$, the curve for the results without the HF term in Fig.\
\ref{Fig3} (open circles) appears even flatter than displayed, which
underlines the importance of this term.
\end{thebibliography}
\end{document}